\documentclass[prl,aps,reprint,amsmath,amssymb,superscriptaddress]
{revtex4-2}
\usepackage{graphicx}% Include figure files
\usepackage{dcolumn}% Align table columns on decimal point
\usepackage{bm}% bold math
\usepackage{color} % For blue in-text comments and additions
\usepackage{physics}
\usepackage{hyperref}
\usepackage{subfigure}
\hypersetup{
    colorlinks = true,
    linkcolor = blue,
    citecolor = blue,
    anchorcolor = blue,
    urlcolor = blue
    }

\def\lesssim{\ \raise.3ex\hbox{$<$}\kern-0.8em\lower.7ex\hbox{$\sim$}\ }
\def\gesim{\ \raise.3ex\hbox{$>$}\kern-0.8em\lower.7ex\hbox{$\sim$}\ }
\def\up{\uparrow}
\def\dwn{\downarrow}

\newcommand \beq{\begin{eqnarray}}
\newcommand \eeq{\end{eqnarray}}

\usepackage{amsmath,amssymb}
\usepackage{fixmath}
\usepackage[normalem]{ulem}

\usepackage{comment}

\begin{document}

\title{Magnonic spin-current shot noise in an itinerant Fermi gas}

\author{Tingyu Zhang}
\affiliation{Department of Physics, School of Science, The University of Tokyo, Tokyo 113-0033, Japan}
\affiliation{Interdisciplinary Theoretical and Mathematical Sciences Program (iTHEMS), RIKEN, Wako, Saitama 351-0198, Japan}
\author{Hiroyuki Tajima}
\affiliation{Department of Physics, School of Science, The University of Tokyo, Tokyo 113-0033, Japan}
\author{Haozhao Liang}
\affiliation{Department of Physics, School of Science, The University of Tokyo, Tokyo 113-0033, Japan}
\affiliation{Interdisciplinary Theoretical and Mathematical Sciences Program (iTHEMS), RIKEN, Wako, Saitama 351-0198, Japan}

\begin{abstract}
{Spin transport phenomena at strongly-correlated interfaces play central roles in fundamental physics as well as spintronic applications.
To anatomize spin-transport carriers, we propose the detection of the spin current noise in interacting itinerant fermions. The Fano factor given by the ratio between the spin current and its noise reflects elementary carriers of spin transport at the interface of spin-polarized Fermi gases realized in ultracold atoms. 
The change of the Fano factor microscopically evinces a crossover from the quasiparticle transport to magnon transport in itinerant fermionic systems.}

\end{abstract}

\maketitle

%\section{Introduction}
\textit{Introduction}.---
Quantum transport phenomena have significantly advanced modern physics, contributing to notable discoveries such as superconductivity, superfluidity, Josephson effect, and quantum Hall effect~\cite{PhysRevB.56.892,sukhatme2001observation,cage2012quantum}.  
Recently, they've been extensively studied in cold atomic systems due to their high controllability and cleanliness. By utilizing Feshbach resonances~\cite{RevModPhys.82.1225}, we are permitted to scan the systems from weakly-interacting to strongly-interacting regimes in both attractive and repulsive branches, enabling the study of transport phenomena over a wide range of interaction strengths~\cite{PhysRevLett.92.040403,PhysRevLett.92.120401}. State-of-the-art experimental techniques allow us to examine various transport phenomena, such as multiple Andreev reflections~\cite{science.aac9584} and Josephson effect~\cite{science350,Krinner_2017,annurev-conmatphys-031218-013732,science.aaz2463} in ultracold atomic systems. 

Spin transport, among these phenomena, has garnered attention in various systems~\cite{PhysRevLett.90.166602,linder2015superconducting,PhysRevB.96.134412,PhysRevB.99.144411,jepsen2020spin,han2020spin,PhysRevB.102.144521}. 
Spin tunneling process at the interface has been investigated in the field of spintronics through the spin Seebeck effect~\cite{uchida2008observation,uchida2010spin,PhysRevB.81.214418,PhysRevB.83.094410,Adachi_2013} and microwave irradiation caused spin pumping effect~\cite{PhysRevLett.88.117601,PhysRevLett.90.166602,kajiwara2010transmission,PhysRevB.89.174417}.
Moreover, spin transport has been studied in ultracold Fermi gases~\cite{Bruun_2011,PhysRevA.88.033630,science.1247425}, offering a controllable interaction strength unlike solid-state systems. The spin current can be directly induced and measured with a spin bias that can be tuned through an external magnetic field~\cite{sommer2011universal,pnas.1601812113},
facilitating the study of spin tunneling phenomena between Fermi gases with a simple two-terminal setup~\cite{PhysRevLett.118.105303,PhysRevResearch.2.023152}.

\begin{figure}[t]
    \centering
    \includegraphics[width=8.5cm]{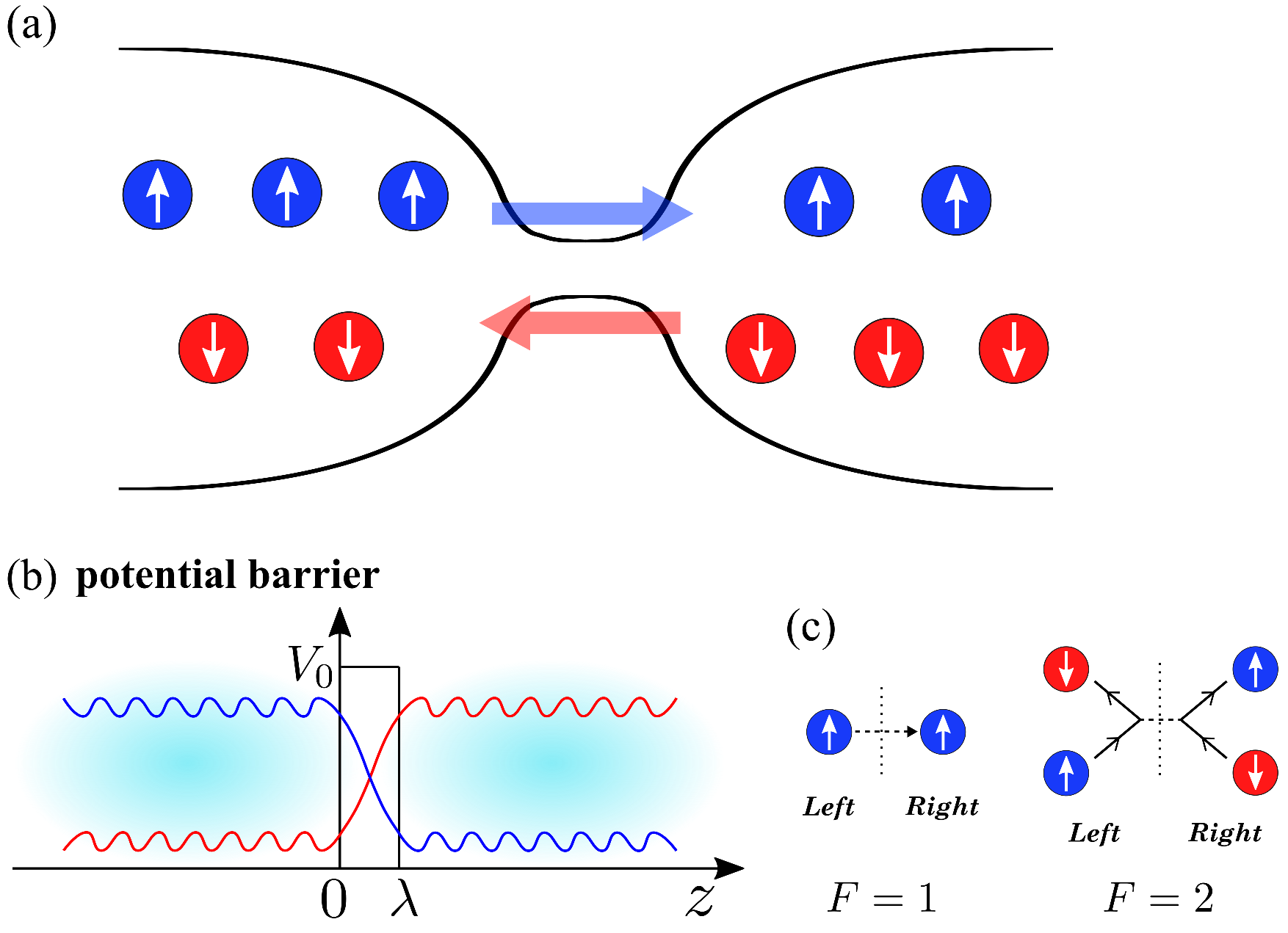}
    \caption{(a) Schematic view of the two-terminal system considered in this work. Spin currents are induced through the potential barrier due to the spin imbalance between two reservoirs. (b) Rectangular potential barrier with height $V_0$ and width $\lambda$ applied for the junction. The blue (red) wavy lines denotes the wave functions of particles in the left (right) reservoir. (c) Quasiparticle tunneling and spin-flip tunneling. At large polarization bias, the Fano factors for these two processes read one and two, respectively.
    }\label{schematic}
\end{figure}

On the other hand, a key mechanism of spin transport, i.e., spin-flip tunneling process at interfaces, remains elusive in ultracold Fermi gases.
In this regard, the tunneling spin transport has been studied based on the conventional quasiparticle tunneling process, unlike ferromagnetic or antiferromagnetic insulators where the quasiparticle tunneling is absent and the spin-flip tunneling process is definitely important~\cite{PhysRevB.83.094410,qiu2016spin}.
It is thus highly required to distinguish elementary spin-transport carriers
for the identification of the spin-flip process in ultracold Fermi gases.
One notable feature of the spin-flip tunneling is the magnon tunneling transport, linked to ferromagnetic phase transitions in repulsive Fermi gases.~\cite{Sandri_2011}.
Recently, it is reported that the spin-flip current exhibits a predominantly cubic dependence on the polarization bias due to the magnon mode~\cite{PhysRevB.108.155303}.
A non-linear conductance associated with the magnon mode has also been observed in magnetic tunneling junctions~\cite{PhysRevB.77.014440,PhysRevB.103.245427,PhysRevB.107.094436}.
While the ferromagnetic phase in the so-called excited branch is metastable and thus exhibits a short lifetime~\cite{PhysRevLett.129.203402}, the spin tunneling transport of the gapped magnon mode still plays a crucial role even in the normal phase.
In particular, a physical observable characterizing a crossover from the quasiparticle to magnon tunneling transport would aid in understanding spin transport in itinerant fermionic systems comprehensively, potentially connecting itinerant and insulating systems in terms of spin transport.

The shot noise~\cite{BLANTER20001} arising from the discrete nature of current carriers is expected to be crucial for studying this issue. Proportional to both the effective charge and the average current, the Fano factor, defined as the noise-to-current ratio, serves as a powerful tool to probe the current carrier. It has been used to detect the effective charge through a normal metal-superconductor junction~\cite{jehl2000detection,PhysRevLett.84.3398}, and to determine the fractional charges in the fractional quantum Hall regime~\cite{de1998direct,PhysRevLett.79.2526}. The Fano factor has also been applied to assess the effective spin carried by a magnon in the spin Seebeck effect and spin pumping induced spin transport~\cite{PhysRevLett.120.037201}.
The spin current noise has been measured at a ferromagnet-metal interface in an electric charge channel via the inverse spin Hall effect~\cite{PhysRevLett.111.066602}, while it is recently proposed that the Fano factor for tunneling in strongly correlated Fermi gases changes from one to two as the interaction strengthens, indicating a crossover from quasiparticle tunneling to pair tunneling dominance~\cite{pgad045}. 
While the magnon-mediated spin current shot noise is found to be small in recent experiments for solid state systems such as ferromagnet-metal junction~\cite{PhysRevB.108.144420}, it can be enhanced by tuning the interaction strength or the tunneling barrier in cold atomic systems.
These findings motivate us to investigate the Fano factor for spin tunneling in spin-polarized Fermi gases.

In this work, to tackle the spin transport at strongly-correlated interface of itinerant systems, we propose a two-terminal junction inducing the spin tunneling through a rectangular potential barrier between two polarized Fermi gases with a repulsive interaction as shown in Figs.~\ref{schematic}(a) and \ref{schematic}(b).
Using the Schwinger-Keldysh approach~\cite{Schwinger,Keldysh}, we analyze the spin current and corresponding shot noise. In large bias limit, 
the Fano factor $F$ can be regarded as a probe of spin carrier 
at large polarization bias (Fig.~\ref{schematic}(c)). We show the interaction and barrier-shape dependencies of $F$, which reveal how the tunneling channel changes from the quasiparticle-dominant to magnon-dominant processes.
Once the spin shot noise is measured in itinerant Fermi gases, the existence of spin-flip tunneling can be clearly revealed
and an analog quantum simulation of spintronics devices can be feasible in systematic ways.

\textit{Model}.---
Throughout the paper, we take $\hbar=k_B=1$ and the volumes for both reservoirs to be unity.
We consider a two-terminal model for two-component Fermi gases with a contact-type repulsive interaction. Despite the potential barrier in the center of the junction, we assume homogeneous gases in the reservoirs far from the barrier. The total Hamiltonian is given by $H=H_{\rm L}+H_{\rm R}+H_{\rm t}$, where the reservoir Hamiltonian reads
\begin{align}\label{Hi}
    &~H_{i={\rm L}, {\rm R}}=\sum_{\bm{p},\sigma}\varepsilon_{\bm{p},\sigma,i}c^\dagger_{\bm{p},
    \sigma,i}c_{\bm{p},\sigma,i}\nonumber\\
    &~+g\sum_{\bm{p},\bm{p}',\bm{P}}c^\dagger_{\bm{p}+\frac{\bm{P}}{2},\uparrow,i}c^\dagger_{-\bm{p}+\frac{\bm{P}}{2},\downarrow,i}c_{-\bm{p}'+\frac{\bm{P}}{2},\downarrow,i}c_{\bm{p}'+\frac{\bm{P}}{2},\uparrow,i}.
\end{align}
Here, $\varepsilon_{\bm{p},\sigma,i}= \epsilon_p - \mu_{\sigma,i} = p^2/(2m)-\mu_{\sigma,i}$ is the single-particle energy in the reservoir $i={\rm L,R}$ with the reservoir chemical potential $\mu_{\sigma,i}$, $g$ is the interaction strength, and $c^\dagger_{\bm{p},\sigma,i}$ ($c_{\bm{p},\sigma,i}$) denotes the single-particle creation (annihilation) operator in the reservoir $i$. For the tunneling Hamiltonian $H_{\rm t}$, we propose an effective Hamiltonian where the pair tunneling is omitted, as we consider zero chemical bias and thus the mass tunneling does not arise. Accordingly, $H_{\rm t}$ includes only the quasiparticle and spin-flip tunneling terms, respectively given by
\begin{align}
    H_{\rm 1t}&=\sum_{\bm{p},\bm{q},\sigma}
    \mathcal{T}_{1,\bm{p},\bm{q}'}
    c^\dagger_{\bm{p},\sigma,{\rm R}}c_{\bm{q},\sigma,{\rm L}}+\rm{H.c.}, \label{H1t} \\
    H_{\rm 2t}&=\sum_{\bm{p},\bm{q}}
    \mathcal{T}_{2,\bm{p},\bm{q}}
    \big(S^+_{\bm{p},{\rm L}}S^-_{\bm{q},{\rm R}}+S^+_{\bm{q},{\rm R}}S^-_{\bm{p},{\rm L}}\big). \label{H2t}
\end{align}
with the spin ladder operators $S^+_{\bm{p},i}=c^\dagger_{\bm{p},\uparrow,i}c_{\bm{p},\downarrow,i}$ and $S^-_{\bm{p},i}=c^\dagger_{\bm{p},\downarrow,i}c_{\bm{p},\uparrow,i}$. Here, $\mathcal{T}_{1,\bm{p},\bm{q}}$ and $\mathcal{T}_{2,\bm{p},\bm{q}}$ respectively represent the tunneling strengths of one-body and spin-flip tunnelings, which depend on the single-particle transmission coefficient~\cite{PhysRevA.106.033310} and hence can be tuned via the potential barrier. 

In this study, we introduce a rectangular barrier with width $\lambda$ and 
height $V_0$: $V(z)=0$ for $z < 0$ and $z > \lambda$, and $V(z)=V_0$ for $0 \leq z \leq \lambda$, as shown in Fig.~\ref{schematic}(b).
Assuming the momentum conservation during the tunneling process 
($\bm{k}-\bm{k}'\rightarrow0$), the tunneling strengths can be obtained from the 
derivation of the tunneling Hamiltonian~\cite{PhysRevA.106.033310} as 
$\mathcal{T}_{1,\bm{k},\bm{k}'}\equiv\mathcal{T}_{1,\bm{k}}\delta_{\bm{k}\bm{k}'}=[C_{\bm{k},\sigma}\epsilon_k+V_0\mathcal{B}_{k_z}]\delta_{\bm{k}\bm{k}'}$ and $\mathcal{T}_{2,\bm{k},\bm{k}'}\equiv\mathcal{T}_{2,\bm{k}}\delta_{\bm{k}\bm{k}'}=2g\operatorname{Re}[C^*_{\bm{k},\up}C_{\bm{k},\dwn}]\delta_{\bm{k}\bm{k}'}$, 
where $C_{\bm{k},\sigma}$ is the transmission coefficient for a single-particle 
tunneling through the barrier with momentum $\bm{k}$ and spin $\sigma$ and 
$\mathcal{B}_{k_z}$ is an overlap integral for wave functions within the potential 
barrier (see the supplemental material). Notice that we can apply such 
momentum-conservation approximation since the contribution of momentum-unconserved 
tunneling to the total current can be much less than the momentum-conserved one.

\begin{figure}[t]
    \centering
    \includegraphics[width=8.6cm]{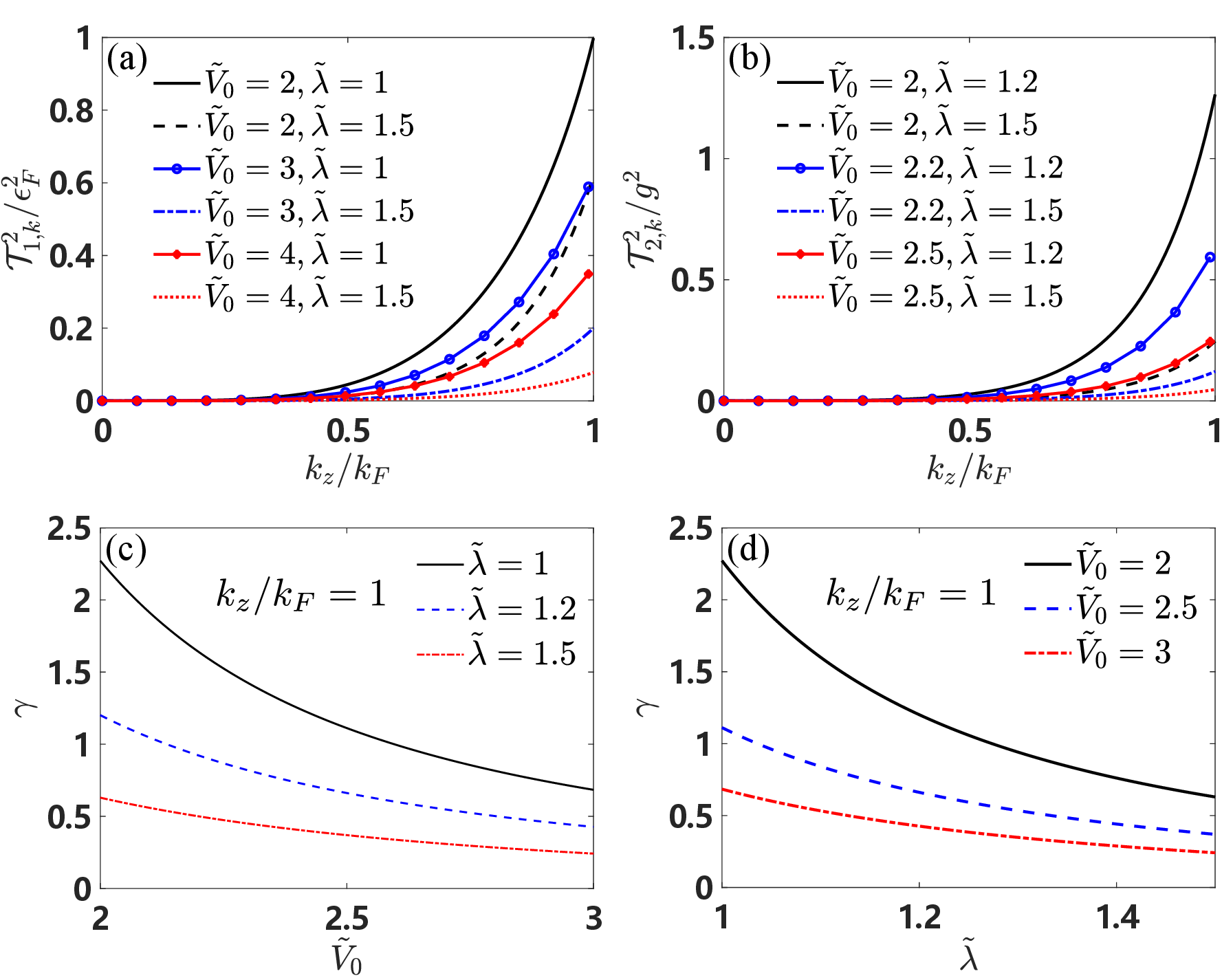}
    \caption{
    Upper panels: Momentum dependence of (a) module squared quasiparticle tunneling strength and (b) module squared spin-flip tunneling strength.
    Lower panels: The ratio $\gamma$ as a function of (c) the height and (d) the width of the potential barrier.
    Here, $\Tilde{V}_0=V_0/\epsilon_{\rm F}$ and $\Tilde{\lambda}=\lambda k_{\rm F}$, and $k_z$ denotes the momentum vector parallel to the axis of the potential barrier.
    }\label{strengths}
\end{figure}

With calculations detailed in the Supplemental Material, the momentum dependence of the tunneling strengths is depicted in Figs.~\ref{strengths}(a) and \ref{strengths}(b), where different features of potential barriers are considered. 
The tunneling of particles near the Fermi surface with large momentum $k_z$ plays a dominant role in the total tunneling process, making it reasonable to approximately regard the tunneling strength as constants 
$\mathcal{T}_1\equiv \mathcal{T}_{1,k_z=k_{\rm F}}$ and $\mathcal{T}_2=\mathcal{T}_{2,k_z=k_{\rm F}}$. In addition, $\mathcal{T}_{2,\bm{k}}$ is shown to be more sensitive than $\mathcal{T}_{1,\bm{k}}$ to the potential barrier, enabling adjustment of the ratio of the two tunneling strengths by barrier tuning.
In this case, we can set $|2\epsilon_{\rm F}\operatorname{Re}
[C^*_{\bm{k},\uparrow}C_{\bm{k},\downarrow}]|/|C_{\bm{k}}\epsilon_{\rm F}+V_0\mathcal{B}_{k_z}|=\gamma$. Figures~\ref{strengths}(c) and \ref{strengths}(d) present how this ratio is affected by the height and width of the potential barrier, respectively.

\textit{Spin current and noise}.---
To calculate the tunneling currents, we apply the mean-field approximation, which yields a reservoir Hamiltonian as 
%\begin{equation}
    $H_i\simeq \sum_{\bm{p},\sigma}\xi^\sigma_{\bm{p},i}c^\dagger_{\bm{p},\sigma,i}c_{\bm{p},\sigma,i}-gn_{\up,i}n_{\dwn,i}$,
%\end{equation}
where we introduced $\xi^\sigma_{\bm{p},i}=\bm{p}^2/2m-\mu'_{\sigma,i}$ with an effective chemical potential $\mu'_{\sigma,i}=\mu_{\sigma,i}-gn_{\bar{\sigma},i}$, while $n_{\sigma,i}=\sum_{\bm{p}}\langle c^\dagger_{\bm{p},\sigma,i}c_{\bm{p},\sigma,i}\rangle$ denotes the particle number density for each component.

The spin current operator is defined as $\hat{I}_{\rm s}=i[N_{\uparrow,{\rm L}},H_{\rm t}]-i[N_{\downarrow,{\rm L}},H_{\rm t}]$, where $N_{\sigma,i}=\sum_{\bm{p}}c^\dagger_{\bm{p},\sigma,i}c_{\bm{p},\sigma,i}$ denotes the particle number of $\sigma$-spin component in reservoir $i$. 
Assuming a nonequilibrium steady state where local equilibrium is permitted in each reservoir, and employing the Schwinger-Keldysh approach with the lowest-order perturbation of $H_{\rm t}$, the expectation value of spin current is obtained as $I_{\rm s}=I_{\rm 1s}+I_{\rm 2s}$. The one-body tunneling contribution reads
\begin{align}\label{I1s}
    I_{\rm 1s}=4&\mathcal{T}_1^2\sum_{\bm{p},\bm{q},\sigma}\int\frac{d\omega}
    {2\pi}\beta_\sigma\big[\operatorname{Im}G^{\rm ret.}_{\bm{q},\sigma,{\rm L}}
    (\omega-\Delta\mu'_{\sigma})\nonumber\\
    &\times\operatorname{Im}G^{\rm ret.}_{\bm{p},\sigma,{\rm R}}(\omega)\big]
    \big[f(\omega-\Delta\mu'_\sigma)-f(\omega)\big],
\end{align}
with the factor $\beta_\up=1$, $\beta_\dwn=-1$ and $\Delta\mu'_\sigma=\mu'_{\sigma,{\rm L}}-\mu'_{\sigma,{\rm R}}$. $G^{\rm ret.}_{\bm{p},\sigma,i}(\omega)$ is the retarded single-particle Green's function in frequency representation and $f(\omega)=1/(e^{\omega/T}+1)$ is the Fermi distribution function. 
Introducing a fictitious magnetic field $h_i=\mu_{\up,i}-\mu_{\dwn,i}$ in each reservoir, 
we obtain the spin-flip tunneling current as  
\begin{align}\label{I2s}
    I_{\rm 2s}=8\mathcal{T}_2^2\sum_{\bm{p},\bm{q}}\int\frac{d\omega}{2\pi}
    &\operatorname{Im}\chi^{\rm ret.}_{\bm{p},{\rm L}}(\omega) \operatorname{Im}
    \chi^{\rm ret.}_{\bm{q},{\rm R}}(\omega-2\Delta h)\nonumber\\
    &\times[b(\omega-2\Delta h)-b(\omega)],
\end{align}
where the polarization bias reads $\Delta h=h_{\rm L}-h_{\rm R}$, $\chi^{\rm ret.}_{\bm{p},i}(\omega)$ is the dynamical spin-flip susceptibility, and $b(\omega)=1/(e^{\omega/T}-1)$ is the Bose distribution function.
In linear response theory, the spin-flip susceptibility is given as a retarded correlation function: $\chi^{\rm ret.}_{\bm{p},i}(t,t')=-i\theta(t-t')\langle[S^+_{\bm{p},i}(t),S^-_{\bm{p},i}(t')]\rangle$, where the quantum average $\langle\cdots\rangle$ represents the expectation value with respect to the unperturbed state of the system and $[S^+_{\bm{p},i}(t),S^-_{\bm{p},i}(t')]=S^+_{\bm{p},i}(t)S^-_{\bm{p},i}(t')-S^-_{\bm{p},i}(t')S^+_{\bm{p},i}(t)$. By applying the random phase approximation~\cite{JPSJ.18.1025,ENGLERT1964429}, the Fourier transform of the spin-flip susceptibility is expressed as $\chi^{\rm ret.}_{\bm{p},i}(\omega)=\Pi_{\bm{p},i}(\omega)/(1+g\Pi_{\bm{p},i}(\omega))$, where $\Pi_{\bm{p},i}(\omega)$ is the Lindhard function~\cite{lindhard}. 

We are now ready to analyze the spin current shot noise $\mathcal{S}$ defined as $\mathcal{S}(t_1,t_2)=\frac{1}{2}\langle\{\hat{I}(t_1)\hat{I}(t_2)\}\rangle$~\cite{PhysRevB.46.12485,BLANTER20001,lumbroso2018electronic}, where $\{\hat{I}(t_1)\hat{I}(t_2)\}=\hat{I}(t_1)\hat{I}(t_2)+\hat{I}(t_2)\hat{I}(t_1)$. We apply the direct-current limit for the shot noise $\mathcal{S}:=\mathcal{S}(\omega\rightarrow 0)$, where $S(\omega)$ is the Fourier transform of $\mathcal{S}(t_1,t_2)$:
\begin{equation}
    \mathcal{S}(\omega)=\frac{1}{\tau}\int_{0}^{\tau}dt_1\int_{0}^{\tau}dt_2
    e^{i\omega(t_1-t_2)}\mathcal{S}(t_1,t_2).
\end{equation}
We introduce the typical time interval for the noise measurement, $\tau$, which ought to be sufficiently larger than the transport time scale~\cite{science.1242308} and can thus be taken as $\tau\rightarrow \infty$. Keeping the result up to the leading order, we obtain the spin shot noise as a sum of one-body (quasiparticle) and spin-flip (magnon) contributions $\mathcal{S}=\mathcal{S}_{\rm 1s}+\mathcal{S}_{\rm 2s}$, where each term reads
\begin{widetext}
\begin{align}
    \mathcal{S}_{1s} &=
    4\mathcal{T}_1^2
    \sum_{\bm{p},\bm{q},\sigma}
    %|\mathcal{T}_{1,\bm{p},\bm{q}}|^2
    \int_{-\infty}^{\infty}
    \frac{d\omega}{2\pi}\operatorname{Im}G^{\rm ret.}_{\bm{p},\sigma,L}
    (\omega-\Delta\mu'_{\sigma})\operatorname{Im}G^{\rm ret.}_{\bm{q},\sigma,R}
    (\omega)
    \big[f(\omega-\Delta\mu'_{\sigma})(1-f(\omega))+(1-f(\omega-
    \Delta\mu'_{\sigma}))f(\omega)\big], \label{S1s} \\
    \mathcal{S}_{\rm 2s} &=16
    \mathcal{T}_2^2\sum_{\bm{p},\bm{q}}
    %|\mathcal{T}_{2,\bm{p},\bm{q}}|^2
    \int_{-\infty}^{\infty}
    \frac{d\omega}{2\pi}\operatorname{Im}\chi^{\rm ret.}_{\bm{p},L}(\omega)
    \operatorname{Im}\chi^{\rm ret.}_{\bm{q},R}(\omega-2\Delta h)
    \big[b(\omega)(1+b(\omega-2\Delta h))+(1+b(\omega))b(\omega-2\Delta h)\big]. \label{S2s}
\end{align}
\end{widetext}
We note that the current shot noise is proportional to the effective spin charge of the current carrier, and one can easily prove that at large polarization bias $\mathcal{S}_{\rm 1s}/I_{\rm 1s}=1$ and $\mathcal{S}_{\rm 2s}/I_{\rm 2s}=2$. This indicates that the Fano factor $F=(\mathcal{S}_{\rm 1s}+\mathcal{S}_{\rm 2s})/(I_{\rm 1s}+I_{\rm 1s})$ changes from one to two as the tunneling channel changes from the quasiparticle-dominated to magnon-dominated transports as shown in Fig.~\ref{schematic}(c).

\begin{figure}[t] 
    \centering
    \includegraphics[width=8.6cm]{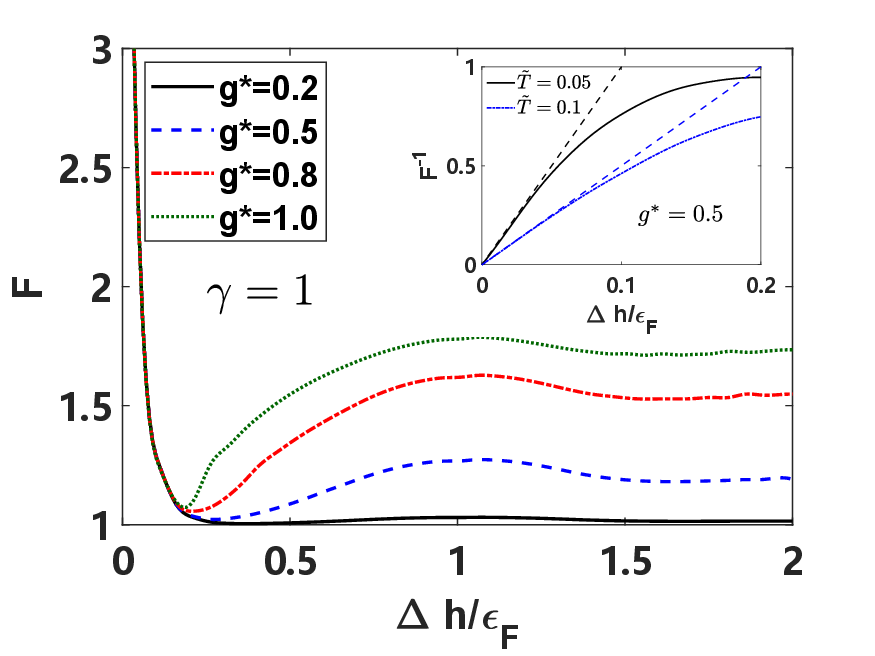}
    \caption{
    Bias dependence of the Fano factor for spin tunneling with different interaction strengths. The dimensionless interaction strength is defined as $g^*=8k_{\rm F}a/3\pi$. The value of $\gamma$ is set to be $1$ and temperature is taken as $\Tilde{T}=T/T_{\rm F}=0.05$. The inset shows the Onsager's relation $F^{-1}(\Delta h\rightarrow 0)=\Delta h/2T$.}
    \label{F-dh}
\end{figure}

\textit{Fano factor for spin transport}.---To investigate the spin tunneling current and the Fano factor, we introduce the dimensionless interaction strength $g^*=g\mathcal{N}/\epsilon_{\rm F}=8k_{\rm F}a/3\pi$ with the scattering length $a$, Fermi momentum $k_{\rm F}$, Fermi energy $\epsilon_{\rm F}$, and $\mathcal{N}=k_{\rm F}^3/3\pi^2$.
Here, $k_{\rm F}$ in both reservoirs are set to be equal for inducing pure spin current.
By fixing the left reservoir to be fully polarized and varying the polarization of the right side, we numerically analyze the $F$-$\Delta h$ characteristics as shown in Fig.~\ref{F-dh}, where the largest bias ($\Delta h/\epsilon_{\rm F}=2$) corresponds to the case of two reservoirs with full but opposite polarization.
This $F$-$\Delta h$ feature indicates that the large-bias limit is valid when $\Delta h$ exceeds approximately $1.5\epsilon_{\rm F}$. In addition, a divergence is found at $\Delta h\rightarrow 0$ due to zero current but finite noise. 
Noticing $\Delta \mu'_{\up}=\Delta h$ and $\Delta \mu'_{\dwn}=-\Delta h$ in Eqs.~(\ref{I1s}) and (\ref{S1s}), as well as the identities $f(\omega)[1-f(\omega)]=-T\partial f(\omega)/\partial \omega$, $b(\omega)[1+b(\omega)]=-T\partial b(\omega)/\partial \omega$, one can get the Onsager's relation near $\Delta h=0$, which yields $F^{-1}=\Delta h/2T$ and is shown in the inset of Fig.~\ref{F-dh}.

\begin{figure}[t]
    \centering
    \includegraphics[width=8.6cm]{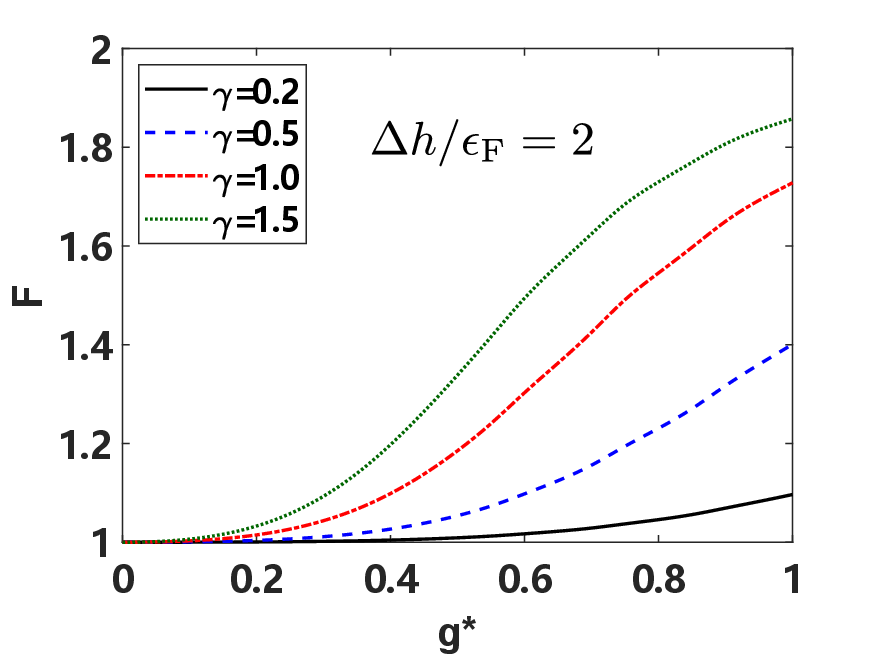}
    \caption{
    Interaction dependence of the Fano factor for spin tunneling with different values of $\gamma$. The Fermi energies of both sides are set to be equal. The polarization bias is taken as $\Delta h/\epsilon_{\rm F}=2$ to maintain the large bias condition and the temperature is set to be $\Tilde{T}=0.05$.}
    \label{F-g}
\end{figure}

Figure~\ref{F-g} demonstrates how $F$ varies with the interaction, where the polarization bias is set as $\Delta h/\epsilon_{\rm F}=2$ to ensure the large-bias limit.
We should note that the repulsive Fermi gas considered here is intrinsically unstable against ground state pair formation~\cite{valtolina2017}, leading to the particle loss via the three-body recombination. To avoid this, the results in Fig.~\ref{F-g} are calculated up to $g^*=1$, which is close to the largest interaction strength reported by experiments~\cite{PhysRevLett.129.203402} for maintaining stability.
At weak coupling ($g^*<0.2$), $F$ remains close to $1$, indicating a dominance of quasiparticle spin tunneling. Beyond that, $F$ increases as the interaction becomes stronger, suggesting an enhanced spin-flip tunneling. For $\gamma=1.5$, the Fano factor reaches a value $F\simeq 1.85$, signifying a magnon-dominated transport for the tunneling processes. 
While the pair-dominant regime is realized in attractively interacting Fermi gases even in the normal phase due to the preformed pair formation~\cite{pgad045}, the dominance of spin-flip tunneling in the present system originates from the stimulation of magnon modes with the strong repulsion~\cite{PhysRevB.32.2824}.

While the spin current has been experimentally measured in cold atomic systems using a quantum point contact~\cite{pnas.1601812113,sommer2011universal}, noise detection experiments have not yet been conducted. However, a theoretical proposal for mass current noise measurement exists~\cite{PhysRevA.98.063619}. In this setup, a single-sided optical cavity overlaps with one reservoir of a two-terminal system, and an incident probe beam monitors the atom number in the reservoir. The current noise can be measured through the phase of the probe. Even in the presence of particle losses, the current noise contains components proportional to the particle loss rate and can thus be experimentally measured~\cite{uchino2023particle}. We anticipate that our theoretical study will spark experimental interest, potentially catalyzing the initiation of noise measurement experiments.
 
%\section{Conclusion}\label{conclusion}
\textit{Conclusion}.---
In this work, we theoretically investigate the spin tunneling current and the associated shot noise induced by a magnetization bias between two repulsively interacting itinerant Fermi gases. We demonstrate that the noise-to-current ratio, known as the Fano factor, can serve as a probe of the current carrier in the spin tunneling process. The Fano factor increases with a stronger repulsive interaction or appropriately tuned potential barrier, indicating a crossover from quasiparticle tunneling to magnon tunneling in itinerant fermions near the ferromagnetic transition. Detecting magnonic transport through the Fano factor and controlling it via engineered potential barriers could significantly advance spin transport development. Our results illustrate magnetic properties in itinerant fermions relevant to repulsive Fermi gases and metals, emphasizing the importance of noise measurement for understanding multi-particle tunneling at strongly correlated interfaces.

\begin{acknowledgements}
The authors thank D.~Oue, M.~Matsuo, T. Kato, and the members in RIKEN iTHEMS NEW working group for useful discussions.
T.Z.~is supported by the RIKEN Junior Research Associate Program.
H.T. is supported by the JSPS Grants-in-Aid for Scientific Research under Grants No.~18H05406, No.~22H01158, and No.~22K13981.
H.L. is supported by the JSPS Grants-in-Aid for Scientific Research under Grant No.~20H05648 and the RIKEN Pioneering Project: Evolution of Matter in the Universe. 
\end{acknowledgements}

\appendix

\section{Supplemental Material: Derivation of tunneling coupling strengths}

In this supplemental material, we derive the tunnel coupling from a microscopic viewpoint.
We consider the one-body barrier potential given by
\begin{align}
    V(z)=\left\{
    \begin{array}{lll}
       V_0,  &  & 0 \leq z \leq \lambda,  \\
       0,  &  & z < 0, \ z > \lambda.
    \end{array}
    \right.
\end{align}
Here, we consider a particle moving from the left reservoir ($z<0$) to the right reservoir ($z > \lambda$) described by $\Psi_{\rm L}(\bm{r})=e^{i\bm{k}_\perp\cdot\bm{r}_\perp}\psi_{{\rm L},z}(z)$, where $\bm{r}_\perp=(x,y)$ and $\bm{k}_\perp=(k_x,k_y)$ are the position and momentum vectors perpendicular to the barrier, respectively. According to the behavior of a wave function passing through a potential barrier, $\psi_{{\rm L}z}(z)$ for a particle coming from the left reservoir can be written as 
\begin{align}
    \psi_{{\rm L},z}(z)=
    \left\{
    \begin{array}{lll}
      e^{ik_z z}+Ae^{-ik_z z}, 
      & & z < 0, \\
        B_{1}e^{\kappa_z z}+B_2e^{-\kappa_z z}, & & 0 \leq z \leq \lambda, \\
    Ce^{ik_z z}, 
    & & z > \lambda.
    \end{array}
    \right.  
\end{align}
where $\kappa_z=\sqrt{2mV_0-k_z^2}$ is a real number (assuming the potential barrier is larger than particle's kinetic energy).
The coefficients $A$, $B_{1,2}$, and $C$ can be obtained by the boundary condition at $z=0$ and $z=\lambda$, where $\psi_z(z)$ and $d\psi_z(z)/dz$ should be continuous. Then, the reflection and transmission coefficients are obtained as 
\begin{align}
    A_{\bm{k}}&=\frac{2(\kappa_z^2+k_z^2)\sinh{(\kappa_z\lambda})}{(\kappa_z+ik_z)^2e^{-\kappa_z\lambda}-(\kappa_z-ik_z)^2e^{\kappa_z\lambda}},\\
    C_{\bm{k}}&=\frac{i4k_z\kappa_ze^{-ik_z\lambda}}{(\kappa_z+ik_z)^2e^{-\kappa_z\lambda}-(\kappa_z-ik_z)^2e^{\kappa_z\lambda}},
\end{align}
while the coefficients $B_1$ and $B_2$ for the wave function inside the potential barrier are obtained as 
\begin{align}
    B_{1,\bm{k}}&=\frac{i2k_z(\kappa_z+ik_z)e^{-\kappa_z\lambda}}{(\kappa_z+ik_z)^2e^{-\kappa_z\lambda}-(\kappa_z-ik_z)^2e^{\kappa_z\lambda}},\\
    B_{2,\bm{k}}&=\frac{i2k_z(\kappa_z-ik_z)e^{\kappa_z\lambda}}{(\kappa_z+ik_z)^2e^{-\kappa_z\lambda}-(\kappa_z-ik_z)^2e^{\kappa_z\lambda}}.
\end{align}
The wave function $\Psi_{{\rm R}}(\bm{r})$ of a particle moving from the right reservoir to the left one can be obtained in a same way.
The total Hamiltonian is now given by
\begin{align}
    H=&~\int d\bm{r}\sum_{\sigma}\hat{\Psi}^\dagger_\sigma(\bm{r})
    \left(-\frac{\nabla^2}{2m}+V(\bm{r})\right)\hat{\Psi}_\sigma(\bm{r})\cr
    &+g\int d\bm{r}\hat{\Psi}_{\uparrow}^\dag(\bm{r})
    \hat{\Psi}_{\downarrow}^\dag(\bm{r})
    \hat{\Psi}_{\downarrow}(\bm{r})
    \hat{\Psi}_{\uparrow}(\bm{r}).
\end{align}
We decompose the fermionic field operator as
\begin{align}
    \hat{\Psi}_{\sigma}(\bm{r})=\hat{\Psi}_{{\rm L},\sigma}(\bm{r})+\hat{\Psi}_{{\rm R},\sigma}(\bm{r}).
\end{align}
Note that we may consider the normalization $\hat{\Psi}_{j,\sigma}(\bm{r})\rightarrow\hat{\Psi}_{j,\sigma}(\bm{r})/\sqrt{2}$ to keep the anti-commutation relations.
However, this procedure does not change the results at all.
We expand $\Psi_{\rm L,R}(\bm{r})$ by using the asymptotic wave functions as
\begin{align}
    \hat{\Psi}_{j={\rm L,R},\sigma}(\bm{r})
    =\sum_{\bm{k}}c_{\bm{k},\sigma,j}\Psi_{j,\sigma}(\bm{r}),
\end{align}
where we only use wave functions $z < 0$ and $z >\lambda$, because we characterize the tunneling strength by using the asymptotic wave functions far away from the tunnel barrier.
Substituting them to the total Hamiltonian and supposing $\bm{k}_\perp$ of the particle to be unchanged before and after tunneling,
the momentum-dependent one-body tunneling strengths are obtained as
\begin{align}
    \mathcal{T}_{1,\bm{k},\bm{k}'}&\simeq C_{\bm{k}'}\delta_{k_zk'_z}\epsilon_{k'}+V_0\mathcal{B}_{k_z,k'_z}.
\end{align}
Here, we omit the Hartree term $gN_{\bm{k}-\bm{k}'}$ which is negligible due to the short-range interaction, while $\mathcal{B}_{k_z,k'_z}=\frac{1}{\lambda}\int_0^\lambda dz\,\psi^\dagger_{{\rm R},z}(z)\psi_{{\rm L},z}(z)$, and $\epsilon_k=k^2/(2m)$ denotes the particle's kinetic energy.
In the limit of $\bm{k}-\bm{k}'\rightarrow 0$, $\mathcal{T}_{1,\bm{k},\bm{k}'}$ can be approximately written as
\begin{equation}
    \mathcal{T}_{1,\bm{k}}\simeq C_{\bm{k}}\epsilon_k+V_0\mathcal{B}_{k_z}.
\end{equation}
Similarly, by applying the momentum-conserved condition,
the spin-flip tunneling strength can be written as
\begin{equation}
    \mathcal{T}_{2,\bm{k}}\simeq 2g\operatorname{Re}[C^*_{\bm{k},\uparrow}C_{\bm{k},\downarrow}],
\end{equation}
where $C_{\bm{k},\sigma}$ is the transmission coefficient for particles with momentum $\bm{k}$ and spin $\sigma$.

\bibliography{ref.bib}

\end{document}